\documentclass[apjl]{emulateapj}
\usepackage{times}
\usepackage{graphicx}
\makeatletter
\begin{document}

\newcommand{\Mo}{M_{\odot}}

\newcommand{\Ro}{R_{\odot}}

\newcommand{\Lo}{L_{\odot}}

\newcommand{\Ms}{M_{\star}}

\newcommand{\Rs}{R_{\star}}

\newcommand{\Ls}{L_{\star}}

\newcommand{\Ts}{T_{\star}}

\newcommand{\rs}{r_{\star}}

\newcommand{\ts}{t_{\star}}

\newcommand{\Ns}{N_{\star}}

\newcommand{\ms}{m_{\bullet}}

\newcommand{\ns}{N_{\bullet}}

\newcommand{\as}{^{\prime\prime}}

\newcommand{\Ua}{\widetilde{a}}

\newcommand{\UP}{\widetilde{P}}

\newcommand{\Um}{\widetilde{m}}

\newcommand{\Ut}{\widetilde{t}}

\newcommand{\Urp}{\widetilde{r}_{p}}

\newcommand{\UdE}{\Delta\widetilde{E}}

\newcommand{\Urt}{\widetilde{r}_{t}}

\newcommand{\Urh}{\widetilde{r}_{h}}

\newcommand{\Urc}{\widetilde{r}_{c}}

\newcommand{\UE}{\widetilde{E}}

\newcommand{\UR}{\widetilde{R}}

\newcommand{\Ue}{\widetilde{\varepsilon}}

\newcommand{\Ums}{\widetilde{m}_{\bullet}}

\newcommand{\Umu}{\widetilde{\mu}}

\newcommand{\Ur}{\widetilde{r}}

\newcommand{\Uv}{\widetilde{v}}

\newcommand{\Pex}{P_{\mathrm{3}}}

\newcommand{\Gs}{\Gamma_{\star}}
\slugcomment{Received 2004 March 4; accepted 2004 March 19; in press}\journalinfo{ApJL, 2004, in press \hfill \today}

\title{Orbital capture of stars by a massive black hole \\
via exchanges with compact remnants}

\author{Tal Alexander}

\affil{Faculty of Physics, Weizmann Institute of Science, P.O. Box 26, Rehovot
76100, Israel; tal.alexander@weizmann.ac.il}

\and

\author{Mario Livio}

\affil{Space Telescope Science Institute, 3700 San Martin Drive, Baltimore
MD 21218; mlivio@stsci.edu }

\begin{abstract}
We propose a dynamical mechanism for capturing stars around a massive
black hole (MBH), which is based on the accumulation of a very dense
cluster of compact stellar remnants near the MBH. This study is motivated
by the presence of $\sim\!10$ young massive stars ($\Ms\!\sim\!3$--$15\,\Mo$,
spectral types $\sim$B9V--O8V) less than $0.04$ pc from the MBH
in the Galactic center (GC). Their existence in the extreme environment
so close to an MBH is a challenge for theories of star formation and
stellar dynamics. We show that young stars, which formed far from
the MBH and were then scattered into eccentric orbits, repeatedly
cross a cluster of stellar black holes (SBHs), where they may undergo
rare direct three-body exchanges with an MBH-SBH {}``binary''. The
interaction between two objects of comparable mass ejects the SBH
and captures the star on a tight orbit around the MBH. Such captures
can naturally explain some trends observed in the orbits of the young
stars. We derive the capture cross-section, validate it by Monte Carlo
simulations, and calculate the number of captured stars in the GC
using the currently uncertain estimates of the numbers of SBHs in
the inner 0.04 pc and of young stars in the inner few parsecs of the
GC. We find that under favorable conditions three-body exchange can
account for $\sim\!25$\% of the observed stars, mostly at the fainter
end of the observed range. We discuss additional effects that possibly
increase the capture efficiency. Future detections of the dark mass
around the MBH and deeper surveys of the central parsecs will establish
whether or not there are enough SBHs and young stars there for exchange
captures to singly account for the central young stars. We estimate
that there are also $\sim\!35$ lower mass stars ($\Ms\!\sim\!1$--$3\,\Mo$,
$\sim$G2V--A0V) in the inner $0.04$ pc similarly captured by exchanges
with neutron stars (NSs). Ongoing replacement of compact remnants
by main-sequence stars (SBHs by NS progenitors, NSs by white dwarf
progenitors) may regulate the accumulation of compact remnants near
the MBH.
\end{abstract}


\section{Introduction}

Deep near-infrared photometric (Krabbe et al. \citeyear{Kra95}; Genzel
et al. \citeyear{Gen03a}), spectroscopic (Genzel et al. \citeyear{Gen97};
Eckart, Ott \& Genzel \citeyear{Eck99}; Figer et al. \citeyear{Fig00};
Gezari et al. \citeyear{Gez02}; Ghez et al. \citeyear{Ghe03a}) and
astrometric (Ghez et al. \citeyear{Ghe03b}; Sch\"odel et al. \citeyear{Sch03})
observations of the dense stellar cusp around the massive black hole
(MBH) in the Galactic center (GC) reveal a centrally concentrated
distribution of young massive stars, $\Ns\sim\!10$ stars within the
central $\rs\sim\!0.04$ pc, $\sim\!40$ within $\sim\!0.1$ pc, spanning
a mass range of $\sim\!3$--$15\,\Mo$ (spectral types $\sim$B9V--O8V)
with a median mass of $\Ms\!\sim\!10\,\Mo$, radius of $\Rs\!\sim\!4.5\,\Ro$
and main sequence life time of $\ts\!\sim\!3\!\times\!10^{7}$ yr%
\footnote{These are rough inferences. The separation of the young stars from
the old population is uncertain since at present only the brightest
star has been spectroscopically identified (Ghez et al. \citeyear{Ghe03a}).%
}.  Orbital solutions obtained for eight of the stars (Ghez et al.
\citeyear{Ghe03b}; Sch\"odel et al. \citeyear{Sch03} ) tentatively
suggest some trends in their orbital properties: a lower bound on
the apoapse of $\sim\!0.01$ pc (Ghez et al. \citeyear{Ghe03b}) and
higher than random orbital eccentricities (Sch\"odel et al. \citeyear{Sch03}).

None of the solutions proposed so far for the puzzle of the young
stars (Genzel et al. \citeyear{Gen03a}; Ghez et al. \citeyear{Ghe03a})
are satisfactory. These fall into three categories: exotic modes of
star formation near the MBH; rejuvenation of old stars in the local
population; or dynamic migration from farther out, where stars can
form. Even if shock cooling of molecular gas by cloud-cloud collisions
could trigger star formation near the MBH (Morris \citeyear{Mor93}),
molecular clouds would either form stars or be tidally disrupted well
outside of the inner 0.1 pc (Vollmer \& Duschl \citeyear{Vol01}).
Growth and rejuvenation by mergers (Genzel et al. \citeyear{Gen03a})
are not expected to be efficient in high velocity collisions near
an MBH. Tidal heating by the MBH requires that the stars approach
the MBH much closer than they are observed to do (Alexander \& Morris
\citeyear{Ale03a}). The young stars are too short-lived to have formed
far from the MBH and then migrated inward by mass segregation or dynamical
friction. The migration can be accelerated if the stars are associated
with a massive {}``anchor'': an extremely dense young cluster (Portegies
Zwart, McMillan \& Gerhard \citeyear{Por03}; Kim \& Morris \citeyear{Kim03}),
a very massive binary companion (Gould \& Quillen \citeyear{Gou03}),
or a $10^{3}$--$10^{4}\,\Mo$ black hole (Hansen \& Milosavljevi\'c
\citeyear{Han03}). However, these scenarios must assume the existence
of very rare, or even hypothetical objects, or else they cannot bring
the stars close enough to the MBH. Such processes may possibly explain
the separate population of very massive and luminous {}``He stars''
$0.1$--$0.5$ pc from the MBH (Krabbe et al. \citeyear{Kra95}),
which we do not attempt to model here. 

Our model is based on the fact that $10^{4}$--$10^{5}$ stellar black
holes (SBHs) of mass $\sim\!7$--$10\,\Mo$ are estimated to exist
within $\sim\!1$ pc of the MBH in the GC, where they have been accumulating
by dynamical friction over the lifetime of the Galaxy ($t_{H}\!\sim\!10\,\mathrm{Gyr}$
) from a {}``collection basin'' $\sim\!10$ pc wide (Morris \citeyear{Mor93};
Miralda-Escud\'{e} \& Gould \citeyear{Mir00}). Numerical simulations
of the evolution of the GC (Freitag \citeyear{Fre03}) confirm that
the SBHs sink to the center on a short timescale of a few gigayears,
settle into a centrally concentrated distribution where the enclosed
number scales as $\ns(<\! r)\!\propto r^{5/4}$ (Bahcall \& Wolf \citeyear{Bah77}),
and dominate the stellar mass there.

\section{Stars and remnants in the Galactic center}

On the $\sim\!1$ pc scale, conditions are more favorable for star
formation. Observations (Figer et al. \citeyear{Fig99}) and theoretical
arguments (Morris \citeyear{Mor93}) indicate that star formation
in the GC is ongoing and is significantly biased toward massive stars.
Here we represent the present-day mass function (PMF) of the central
few parsecs of the GC by a simple (nonunique) model that has such
properties. We assume that the PMF is the product of continuous star
formation at a constant rate with a Salpeter initial mass function
(IMF) in the range $\Ms\!=\!1.5$--$120\,\Mo$, so that $\mathrm{d}\Ns/\mathrm{d}\Ms(t)\!\propto\!\Ms^{-2.35}\min[t,\ts(\Ms)]$.
Stars with $\Ms\!\le\!8\,\Mo$ are assumed to evolve into $0.6\,\Mo$
white dwarfs; those with $\Ms\!=\!8$--$30\,\Mo$ into $1.4\,\Mo$
neutron stars (NSs); and those with $\Ms\!>\!30\,\Mo$ into $7\,\Mo$
SBHs (consistent with the mass distribution found
in black hole binaries; McClintock \& Remillard \citeyear{McC03}).
The gas lost in the course of stellar evolution is assumed to be efficiently
expelled from the system. Using the Schaller et al. (\citeyear{Sch92})
solar metallicity stellar evolution tracks, we find that at $t\!=\! t_{H}\!\sim\!10$
Gyr, the mass fraction in stars out of the total mass is $0.22$,
the mean stellar mass is $2\,\Mo$ and the number fraction of young
stars in the range $3$--$15\,\Mo$ is $f_{\star}\!=\!0.06$, with
a mean mass of $\bar{M}_{\star}\!=\!4\,\Mo$ (corresponding to $\Rs\!=\!2.34\,\Ro$
and $\ts\!=\!2\!\times\!10^{8}$ yr). Dynamical measurements of the
central gravitational potential indicate that the MBH mass is $m\sim\!3\!\times\!10^{6}\,\Mo$
(Sch\"odel et al. \citeyear{Sch03}; Ghez et al. \citeyear{Ghe03b})
and that in the range $\sim\!0.5$--$10$ pc the stellar mass distribution
is well represented by a spherically symmetric power-law distribution,
$\rho_{\star}\!\propto r^{-\alpha}$, with the most
recent calibration giving $\rho_{\star}(1\,\mathrm{pc})\!=\!1.2\!\times\!10^{5}\,\Mo\,\mathrm{pc^{-3}}$
and $\alpha=1.7$ (R. Genzel 2003, private communication). The mass
distribution and mass function translate to $1.5\!\times\!10^{5}$
SBHs born inside the collection basin ($\le\!5$ pc) and to $2.3\!\times\!10^{4}$
young stars inside $2.5$ pc, where the enclosed dynamical mass is
$3.4\!\times\!10^{6}\,\Mo$, in general agreement with the observationally
based PMF in the inner $15$ pc of Mezger et al. (\citeyear{Mez99}),
scaled to $2.5$ pc. 

Young stars formed on the $\gtrsim\!1$ pc scale will not be on eccentric
orbits initially, because their progenitor clouds cannot survive the
tidal field of the MBH. Over time, the stars will be scattered by
gravitational perturbations. The most efficient of these, which can
operate on a timescale of less than $\ts$, are due to massive star-forming
clusters (Zhao, Haehnelt \& Rees \citeyear{Zha02}). One such cluster
is statistically expected to exist, undetected, within a few parsecs
of the MBH (Portegies Zwart et al. \citeyear{Por02}). It is also
possible, albeit speculative, that cloud-cloud collisions on the few
parsecs scale could lead to rapid formation of massive stars on radial
orbits.

Genzel et al. (\citeyear{Gen03a}) estimate that the enclosed \emph{stellar}
mass inside $r\!=\!0.4$ pc scales as $\Ms(<\! r)\!=\!2.5\!\times\!10^{6}(r/1\,\mathrm{pc})^{1.63}\,\Mo$,
assuming that the mass follows the star counts from $\sim\!1$ pc
to the center. However, the \emph{total} mass distribution (stars
and compact remnants) inside $0.1$ pc, where the MBH dominates the
potential but where the remnants dominate the extended mass, is not
known (Mouawad et al. \citeyear{Mou04}). The central density of the
SBH cluster depends on various uncertain quantities: the SBH mass
function, the stellar IMF and formation rate, the remnant progenitor
masses, and the dynamical age of the GC. A rough upper limit \emph{}on
$\ns(<\!\rs)$ can be obtained by requiring that the SBHs survive
being drained into the MBH over the lifetime of the system, $\mathrm{d}\ns/\mathrm{d}t\!<\!\ns/t_{H}$,
where $\mathrm{d}\ns/\mathrm{d}t\!\sim\!\ns/[\log(2/\vartheta_{\mathrm{lc}})t_{r}]$
is the scattering rate into loss-cone orbits that take an SBH into
the MBH, $t_{r}\!\sim\! v^{3}/(\log\Lambda G^{2}\rho\ms)$ is the
relaxation time (assuming all the mass is in SBHs), $\vartheta_{\mathrm{lc}}$
is the loss-cone opening angle and $\Lambda\!\sim\!0.4\ns$ is the
Coulomb cutoff (e.g., Syer \& Ulmer \citeyear{Sye99}). This constraint
yields the relation (solved numerically), 

\begin{equation}
\max\ns(<\rs)\!\sim\!\frac{2\log(2\sqrt{\rs/r_{s}})}{3\log(0.4\max\ns)}\left(\frac{m}{m_{\bullet}}\right)^{2}\frac{P(\rs)}{t_{H}}\,,\label{eq:maxN}\end{equation}
where $r_{s}$ is the Schwarzschild radius of the MBH and $P(\rs)$
is the orbital period at $\rs$.

\section{The direct exchange cross section}

We estimate the efficiency of the capture mechanism with the assumption
that the angular momentum distribution of the young stars has been
efficiently randomized. For simplicity, orbital periods and periapses
are considered Keplerian (this is reasonable inside $2.5$ pc, where
the enclosed mass is $\lesssim\! m$) and the typical orbital energy
of a star is represented by its virial energy (equivalent to assuming
the star was initially on a circular orbit). If the star is scattered
to an eccentric orbit, it will pass by the MBH on a hyperbolic orbit
relative to it, with energy (Alexander \& Livio \citeyear{Ale01b})\begin{equation}
\UE_{0}=+\frac{\Um}{\Ua_{0}}\left[(\mu_{0}\!-\!1)\frac{3\!-\!\alpha}{2\!-\!\alpha}\!-\!\frac{\mu_{0}}{2}\right]\!\lesssim\!0.1\,,\label{eq:E0}\end{equation}
 where $a_{0}$ is the orbital semi-major axis (SMA) and $\mu_{0}m$
is the total mass enclosed within $a_{0}$ ( the tilde symbol denotes
quantities in dimensionless units $G$=$\Ms$=$\Rs$=$1$; in these
units, the stellar binding energy is $\sim\!1$). 

A three-body encounter (Heggie \citeyear{Heg75}) is characterized
by a dimensionless relative velocity $\widehat{v}_{\infty}^{2}\!\equiv\!-\UE_{0}/\UE_{\bullet}$
between the incoming star at infinity and the binary barycenter, where
$\UE_{\bullet}=-\Um\Ums/2\Ua_{\bullet}$ is the binding energy of
an MBH-SBH {}``binary'' with SMA $\Ua_{\bullet}$. When $\widehat{v}_{\infty}\!\ge\!1$
(a fast encounter with a soft binary), the incoming star carries enough
energy to ionize (disrupt) the binary. When $\widehat{v}_{\infty}\!<\!1$
(a slow encounter with a hard binary), the only possible outcomes
are exchange, where the incoming star ejects one of the binary members
and replaces it, or scattering, where the incoming star remains unbound.
Typical three-body encounters between an $\sim\!10\,\Mo$ star and
a SBH in the GC occur below the ionization threshold (Table \ref{t:XGC}).
Note that the actual 2-body interaction between the star and the SBH
is analogous to the high velocity exchanges studied by Hut (\citeyear{Hut83}).

\begin{figure}
\includegraphics[%
  scale=0.7]{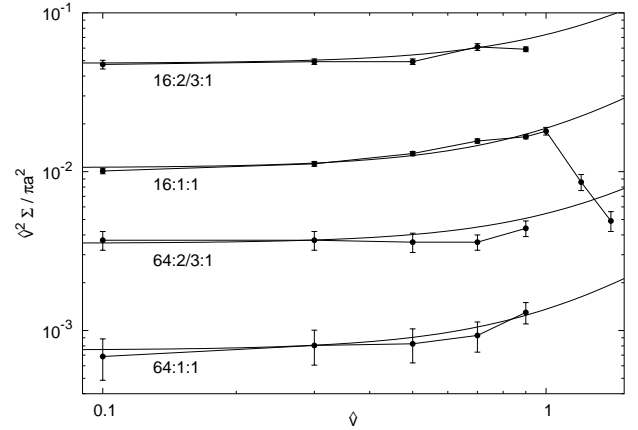}

\figcaption[FileName]{\label{f:X}Direct exchange cross-section below the ionization threshold
(multiplied by $\widehat{v}^{2}/\pi a_{\bullet}^{2}$ ) for $\Um_{\bullet}\!=\!2/3,1$
and $\Um\!=\!16,64$ as a function of $\widehat{v}$. Points with
error bars are Monte Carlo estimates of the cross section (using the
\textsc{sigma3} program {[}McMillan \& Hut \citeyear{McM96}{]} of
the Starlab software package by Hut, McMillan, Makino \& Portegies
Zwart; available at http://www.ids.ias.edu/\textasciitilde starlab).
Thick lines are the best-fit analytic approximation (Eq. \ref{eq:sigma}).
For $\widehat{v}\!>\!1$, binary ionization competes with direct exchange,
and the cross section (shown here only for $m\!:\! m_{\bullet}\!:\!\Ms\!=\!16\!:\!1\!:\!1$)
rapidly falls off.}
\end{figure}

An exchange can proceed via two channels: direct or resonant. It is
direct when the star passes within the SBH capture radius $\Ur_{c}\!\sim[(1+\Ums)/\Um]\Ua_{\bullet}$
and ejects it. In this case, the three-body analogy is justified because
$r_{c}\!\ll\! n_{\star}^{-1/3}$, the mean distance between stars.
It is resonant when a transient three-body bound system forms and
persists for many orbits over a volume of radius $\sim a_{\bullet}$,
until one of the masses is ejected (Hut \citeyear{Hut93a}). In this
case, the isolated three-body analogy is no longer valid because $a_{\bullet}\!\gg\! n_{\star}^{-1/3}$.
The direct exchange cross-section, which is the relevant channel here,
is smaller by a factor of $[(1+\Ums)/\Um]^{2}$ than the resonant
one (Heggie, Hut \& McMillan \citeyear{Heg96}) and was therefore
neglected in past works on isolated three-body encounters. We derive
it using the approximate equality between the binding energy of the
original and exchanged binary (Heggie et al. \citeyear{Heg96}), and
we calibrate it by detailed numerical 3-body simulations (Fig. \ref{f:X}).
In the limit $\Um\gg\Ums,$1 and for $1/2\!\lesssim\!\Ums\!\lesssim\!2$,
the direct exchange cross-section averaged over all orbital angles
for a thermal distribution of binaries with SMA $\Ua_{\bullet}$ is\begin{eqnarray}
\widetilde{\Sigma}(\Ua_{\bullet}) & \simeq & \left(1\!+\!2A\frac{\Um\!+\!\Ums}{\Um\Ums\widehat{v}_{\infty}^{2}}\right)\times\nonumber \\
 &  & \!\!\!\!\!\!\!\!\!\!\!\!\!\!\!\!\!\!\!\!\left[B\pi\left(\frac{1\!+\!\Ums}{1\!+\!\Um\!+\!\Ums}\Ua_{\bullet}\right)^{2}\right]\!\left[\Ums^{-7/4}\left(\frac{\Ums\!+\!\Um}{1\!+\!\Um}\right)^{1/4}\right]\,,\label{eq:sigma}\end{eqnarray}
where the first factor is the gravitational focusing term, the second
is the geometrical term $\pi\Ur_{c}^{2}$, and the third is the phase
space factors required by detailed balance arguments (Heggie et al.
\citeyear{Heg96}). The best fit parameters obtained from the numerical
simulations are $A\!\sim\!0.6$, and $B\!\sim\!0.7$. 

Equation (\ref{eq:sigma}) describes an isolated system in which the
binary has a specified SMA and in which the captured star is initially
unbound at infinity. Capture by three-body exchange in the CG differs
in that the stars are scattered from a finite distance and are captured
by a distribution of SBHs enclosed within $\rs$. Equation (\ref{eq:sigma})
is then applied by replacing $\Ua_{\bullet}\!\rightarrow\left\langle \Ua_{\bullet}\right\rangle \!\sim\!1.5\rs$,
the mean SMA in a $r^{-7/4}$ cusp, and by replacing 

\begin{equation}
\widehat{v}_{\infty}^{2}\!\rightarrow\!\widehat{v}_{0}^{2}\!\left/\!\!\left\{ \!1\!+\!\frac{\rs}{a_{0}}\!\left[\frac{3\!-\!\alpha}{2\!-\!\alpha}(\mu_{0}\!-\!1)\!-\!\mu_{0}\right]\right\} \!,\right.\label{eq:v0}\end{equation}
where $\!\widehat{v}_{0}^{2}\!=\!-(\mu_{0}\Um/\Ua_{0})/(2\UE_{\bullet})$.
The ionization threshold remains at $\widehat{v}_{\infty}\!=\!1$,
but now ionization implies the ejection of the SBH out to a distance
$\sim\! a_{0}$, where its orbit is randomized, and not to infinity.

The number of captured stars orbiting the MBH is the product of the
capture probability per passage, $\ns\Sigma$, the typical lifetime
of a star after capture, $\ts/2$, and the incoming flux of young
stars, $f_{\star}\!\int(n_{\star}q_{\star}/P_{0})\mathrm{d}^{3}r/(\pi r_{\star}^{2})$,
where $P_{0}$ is the orbital period at $a_{0}$ and $q_{\star}\!=\!1\!-\!(1\!-\!\rs/a_{0})^{2}$
is the fraction of stars on orbits that cross inside $\rs$. The steady
state number of captured young stars within $\rs$ is then\begin{equation}
\Ns\!=\!\frac{t_{\star}\ns}{2\pi\rs^{2}}\frac{f_{\star}}{\bar{M}_{\star}}\int_{a_{1}}^{a_{2}}\frac{\nu_{\star}q_{\star}\Sigma}{P_{0}}\,\mathrm{d}a_{0}\,,\label{eq:iNs}\end{equation}
where $\nu_{\star}$ is the stellar mass density per unit SMA.%
\footnote{The spatial distribution $\rho_{\star}(r)\!=\! Cr^{-\alpha}$ in a
Keplerian potential corresponds to $\nu_{\star}(a)\!=\!3\pi2^{-\alpha}\beta(\alpha\!+\!1,-3/2)Ca^{2-\alpha}$
(Sch\"odel et al. \citeyear{Sch03}).%
} The integration runs from the smallest SMA where young stars can
be formed (here assumed to be $a_{1}\!=\!0.44$ pc, where a star with
the virial energy is unbound to the MBH) up to $a_{2}\!=\!2.5$ pc
where the Keplerian approximation is marginally valid. Stars can also
be captured from orbits with $a_{0}\!>\! a_{2}$ up to a maximal SMA,
where $\widehat{v}_{\infty}\!=\!1$. However, the contribution of
such stars to the total is small because of the longer $P_{0}$ and
smaller $q_{\star}$. Note that since $\widehat{v}_{\infty}$ increases
with $a_{\bullet}$ and $a_{0}$, there exists a maximal radius where
stars can be captured, $\rs\!\lesssim\!1$ pc. Beyond this limit three-body
interactions result in ionization, not capture.

\section{Exchange capture in the Galactic Center}

\begin{table}

\caption{\label{t:XGC}Three-body exchange in the Galactic center}

\begin{center}\begin{tabular}{lrr}
\hline 
\multicolumn{3}{r}{\small Extent of young stars, $\rs$ (pc)}\tabularnewline
{\small Parameter\vspace{0.5em}}&
\textsf{\small $0.04$}&
\textsf{\small $0.10$}\tabularnewline
\hline 
\small Number of observed young stars, $\Ns$&
\textsf{\small $\sim\!10$}&
\textsf{\small $\sim\!40$}\tabularnewline
{\small Number of captured young stars in $\rs$},
{\small $\Ns\,^{a}$}&
\textsf{\small $2.4$}&
\textsf{\small $7.7$}\tabularnewline
{\small Number of SBHs in $\rs$, $\ns\,^{b}$}&
\textsf{\small $4.2\!\times\!10^{3}$}&
\textsf{\small $1.3\!\times\!10^{4}$}\tabularnewline
{\small Drain limit, $\max\ns$}&
\textsf{\small $4.7\!\times\!10^{3}$}&
\textsf{\small $1.7\!\times\!10^{4}$}\tabularnewline
{\small Enclosed SBH / stellar mass$^{c}$ ratio in $\rs$ }&
\textsf{\small $2.1$}&
\textsf{\small $1.5$}\tabularnewline
{\small Mean$^{d}$ velocity at infinity, $\left\langle
\hat{v}_{\infty}\right\rangle $}&
{\small $0.2$}&
{\small $0.4$}\tabularnewline
{\small Median$^{e}$ initial SMA, $a_{0}$ (pc)}&
{\small $0.9$}&
{\small $0.9$}\tabularnewline
\hline 
\multicolumn{3}{l}{{\scriptsize $^{a}\,$For $2.3\!\times\!10^{4}$ young stars ($\Mo\!=\!3$--$15\,\Mo$)
inside $2.5$ pc.}}\tabularnewline
\multicolumn{3}{l}{{\scriptsize $^{b}\,$For $1.5\!\times\!10^{5}$ SBHs of $7\,\Mo$
within $r_{\bullet}\!=\!0.7$ pc}\textsf{\scriptsize .}}\tabularnewline
\multicolumn{3}{l}{{\scriptsize $^{c}\,$Relative to the estimated enclosed stellar mass}
\textsf{\scriptsize }{\scriptsize (Genzel et al. \citeyear{Gen03a})}\textsf{\scriptsize .}}\tabularnewline
\multicolumn{3}{l}{{\scriptsize $^{d}\,$$\mathrm{d}\Ns$-weighted mean between $a_{1}$
and $a_{2}$.}}\tabularnewline
\multicolumn{3}{l}{{\scriptsize $^{e}\,$Half of the captured stars originate between
median $a_{0}$ and $a_{2}$. }}\tabularnewline
\hline
\end{tabular}\end{center}

\end{table}

Table (\ref{t:XGC}) shows the number of captured stars, assuming
that SBHs born in the central $5$ pc are now concentrated inside
$r_{\bullet}\!=\!0.7$ pc. This choice of $r_{\bullet}$ (e.g. Morris
\citeyear{Mor93}; Miralda-Escud\'{e} \& Gould \citeyear{Mir00})
implies that the central concentration of SBHs is near the drain limit
(Eq. \ref{eq:maxN}) and that SBHs dominate the central mass density,
as expected in a dynamically evolved system (Freitag \citeyear{Fre03}).
The predicted mean number of captured stars is $\sim\!25$\% of those
observed. The captured stars are expected to lie mostly at the low-mass
and low-luminosity end of the observed range because lower-mass stars
are more numerous and live longer. 

In addition to the fact that capture by exchange can account for at
least some of the young stars, this mechanism can naturally explain
observed trends in their orbital properties. Because $r_{c}$ falls
with the distance to the MBH, there exists a minimal distance $\Ur_{\mathrm{min}}\!\sim\!0.8\Um\Ums^{1/3}/(1+\Ums)$
where the capture radius equals the tidal radius of the SBH (Alexander
\& Kumar \citeyear{Ale01a}). Beyond this limit, an exchange is no
longer possible since the star is disrupted. Near disruption at $r_{\mathrm{min}}$,
the tidal interaction absorbs orbital energy of the order of the stellar
binding energy, so that the captured star loses most of its kinetic
energy in the encounter, and $r_{\mathrm{min}}$ then becomes the
apoapse of the new orbit. In the GC, $r_{\mathrm{min}}\!\sim\!0.01$
pc, which is consistent with the observed lower bound on the apoapse
of the young stars (the local relaxation time is longer than $\ts$,
so they are expected to remain on their original orbits). We note
that the extraction of orbital energy can increase the capture efficiency.
A rough estimate indicates a factor of $\sim\!2$ increase in the
cross-section for captures occurring near $r_{\mathrm{min}}$, where
the tidal interaction is strong. This effect was not included in the
results quoted in Table (\ref{t:XGC}). 

The numerical simulations also show that the exchanged binaries have
higher than random eccentricities, even when the SBHs have a random
(thermal) distribution of eccentricities. The bias becomes substantial
the faster ($\widehat{v}_{\infty}\!\rightarrow\!1$) and more massive
($\Ums\!<\!1$) the incoming star is. This property of the exchange
mechanism agrees with the observed trend. 

Lower mass main-sequence stars in the range $\Ms\!=\!1$--$3\,\Mo$
($\sim$G2V--A0V) can be similarly captured by NSs. Dynamical simulations
of the GC indicate that the central number density of NSs is $\sim1/3$
that of the SBHs (M. Freitag 2003, private communication). We find
that there are $\sim\!35$ lower-mass stars captured in the inner
$0.04$ pc and $\sim\!110$ captured in the inner $0.1$ pc.

\section{Discussion and summary}

The efficiency of a {}``billiard ball'' recoil depends strongly
on the mass ratio of the colliding objects. The dynamical evolution
of a stellar system around an MBH naturally provides a dense concentration
of targets whose masses are well matched for stopping and capturing
unbound young stars like those observed very near the MBH in the GC.
The attempt to predict the number of captured stars is limited by
the uncertainty in the mass and number distribution of the SBHs and
in the number and orbital properties of young stars (spectral types
$\sim$B9V--O8V) in the inner few parsecs of the GC. We show that
under favorable conditions, capture can account for $\sim\!25$\%
of the observed young stars. 

Additional effects that were not taken into account here may increase
the capture efficiency. We considered only capture by a single direct
exchange in the point-mass approximation. However, a few weaker interactions
may also lead gradually to capture during the star's lifetime. Further
study is needed to estimate the contribution of multiple scatterings
to the capture cross section. Tidal energy extraction in captures
near the tidal limit also increases the capture cross section there.
A possible outcome of internal mixing by a strong tidal interaction
is an extended main-sequence lifetime and higher luminosity (Maeder
\& Meynet \citeyear{Mae00}). This would further increase the number
of captured luminous stars (Eq. \ref{eq:iNs}). 

In the context of direct exchange capture, the stars with the smallest
apoapse offer an opportunity to study the long term effects of a strong
tidal interaction. Interestingly, S2, the star with the smallest apoapse,
is also the brightest (Ghez et al. \citeyear{Ghe03b}; Sch\"odel
et al. \citeyear{Sch03}). The orbital eccentricity distribution reflects
the mass ratio between the star and the SBH. Thus, a spectral determination
of the masses of the captured stars together with a statistical analysis
of their eccentricities can probe the poorly known mass function of
SBHs. The total number of captured stars over the lifetime of the
Galaxy (assuming steady state), $\sim\!2\Ns t_{H}/\ts\!\sim\!3\!\times\!10^{4}$
(for $40$ stars of $10\,\Mo$ in 0.1 pc), is of the same order as
the number of SBHs in the inner parsec. Thus, if the dynamical friction
timescale at $\left\langle a_{0}\right\rangle $ (where the SBHs are
ejected to) is not much smaller than the age of the Galaxy, the continual
replacement of SBHs by NS progenitors and of NSs by white dwarf progenitors
(Fryer \& Kalogera \citeyear{Fry01}) may regulate the build-up of
the dense cusp of compact stellar remnants.

\acknowledgements{We thank P. Hut for his generous help and R. Genzel, B. Hansen, D.
Figer and M. Freitag for useful discussions. T.A. is supported by
ISF grant 295/02-1, Minerva grant 8484 and a New Faculty grant by
Sir H. Djangoly, CBE, of London. }

\bibliographystyle{apj}

\begin{thebibliography}{36}
\expandafter\ifx\csname natexlab\endcsname\relax\def\natexlab#1{#1}\fi

\bibitem[{{Alexander} \& {Kumar}(2001)}]{Ale01a}
{Alexander}, T. \& {Kumar}, P. 2001, \apj, 549, 948

\bibitem[{{Alexander} \& {Livio}(2001)}]{Ale01b}
{Alexander}, T. \& {Livio}, M. 2001, \apjl, 560, L143

\bibitem[{{Alexander} \& {Morris}(2003)}]{Ale03a}
{Alexander}, T. \& {Morris}, M. 2003, \apjl, 590, L25

\bibitem[{{Bahcall} \& {Wolf}(1977)}]{Bah77}
{Bahcall}, J.~N. \& {Wolf}, R.~A. 1977, \apj, 216, 883

\bibitem[{{Eckart} {et~al.}(1999){Eckart}, {Ott}, \& {Genzel}}]{Eck99}
{Eckart}, A., {Ott}, T., \& {Genzel}, R. 1999, \aap, 352, L22

\bibitem[{{Figer} {et~al.}(1999){Figer}, {Kim}, {Morris}, {Serabyn}, {Rich}, \&
  {McLean}}]{Fig99}
{Figer}, D.~F., {Kim}, S.~S., {Morris}, M., {Serabyn}, E., {Rich}, R.~M., \&
  {McLean}, I.~S. 1999, \apj, 525, 750

\bibitem[{{Figer} {et~al.}(2000)}]{Fig00}
{Figer}, D.~F. {et~al.} 2000, \apjl, 533, L49

\bibitem[{{Freitag}(2003)}]{Fre03}
{Freitag}, M. 2003, \apjl, 583, L21

\bibitem[{{Fryer} \& {Kalogera}(2001)}]{Fry01}
{Fryer}, C.~L. \& {Kalogera}, V. 2001, \apj, 554, 548

\bibitem[{{Genzel} {et~al.}(1997){Genzel}, {Eckart}, {Ott}, \&
  {Eisenhauer}}]{Gen97}
{Genzel}, R., {Eckart}, A., {Ott}, T., \& {Eisenhauer}, F. 1997, \mnras, 291,
  219

\bibitem[{{Genzel} {et~al.}(2003){Genzel}, {Sch{\" o}del}, {Ott}, {Eisenhauer},
  {Hofmann}, {Lehnert}, {Eckart}, {Alexander}, {Sternberg}, {Lenzen}, {Cl{\'
  e}net}, {Lacombe}, {Rouan}, {Renzini}, \& {Tacconi-Garman}}]{Gen03a}
{Genzel}, R., {Sch{\" o}del}, R., {Ott}, T., {Eisenhauer}, F., {Hofmann}, R.,
  {Lehnert}, M., {Eckart}, A., {Alexander}, T., {Sternberg}, A., {Lenzen}, R.,
  {Cl{\' e}net}, Y., {Lacombe}, F., {Rouan}, D., {Renzini}, A., \&
  {Tacconi-Garman}, L.~E. 2003, \apj, 594, 812

\bibitem[{{Gezari} {et~al.}(2002){Gezari}, {Ghez}, {Becklin}, {Larkin},
  {McLean}, \& {Morris}}]{Gez02}
{Gezari}, S., {Ghez}, A.~M., {Becklin}, E.~E., {Larkin}, J., {McLean}, I.~S.,
  \& {Morris}, M. 2002, \apj, 576, 790

\bibitem[{Ghez {et~al.}(2003)Ghez, Salim, Hornstein, Tanner, Morris, Becklin,
  \& Duchene}]{Ghe03b}
Ghez, A.~M., Salim, S., Hornstein, S.~D., Tanner, A., Morris, M., Becklin,
  E.~E., \& Duchene, G. 2003, {\apj} submitted, available at
  http://xxx.lanl.gov/abs/astro-ph/0306130

\bibitem[{{Ghez} {et~al.}(2003)}]{Ghe03a}
{Ghez}, A.~M. {et~al.} 2003, \apjl, 586, L127

\bibitem[{{Gould} \& {Quillen}(2003)}]{Gou03}
{Gould}, A. \& {Quillen}, A.~C. 2003, \apj, 592, 935

\bibitem[{{Hansen} \& {Milosavljevi{\' c}}(2003)}]{Han03}
{Hansen}, B.~M.~S. \& {Milosavljevi{\' c}}, M. 2003, \apjl, 593, L77

\bibitem[{{Heggie}(1975)}]{Heg75}
{Heggie}, D.~C. 1975, \mnras, 173, 729

\bibitem[{{Heggie} {et~al.}(1996){Heggie}, {Hut}, \& {McMillan}}]{Heg96}
{Heggie}, D.~C., {Hut}, P., \& {McMillan}, S.~L.~W. 1996, \apj, 467, 359

\bibitem[{{Hut}(1983)}]{Hut83}
{Hut}, P. 1983, \apj, 268, 342

\bibitem[{{Hut}(1993)}]{Hut93a}
---. 1993, \apj, 403, 256

\bibitem[{{Kim} \& {Morris}(2003)}]{Kim03}
{Kim}, S.~S. \& {Morris}, M. 2003, \apj, 597, 312

\bibitem[{{Krabbe} {et~al.}(1995)}]{Kra95}
{Krabbe}, A. {et~al.} 1995, \apjl, 447, L95

\bibitem[{{Maeder} \& {Meynet}(2000)}]{Mae00}
{Maeder}, A. \& {Meynet}, G. 2000, \araa, 38, 143

\bibitem[{{McClintock} \& {Remillard}(2004)}]{McC03}
{McClintock}, J.~E. \& {Remillard}, R.~A. 2004, in Compact Stellar X-ray
  Sources, in press, ed. W.~H.~G. Lewin \& M.~van~der Klis (Cambridge:
  Cambridge University Press), available at
  http://xxx.lanl.gov/abs/astro-ph/0306213

\bibitem[{{McMillan} \& {Hut}(1996)}]{McM96}
{McMillan}, S.~L.~W. \& {Hut}, P. 1996, \apj, 467, 348

\bibitem[{{Mezger} {et~al.}(1999){Mezger}, {Zylka}, {Philipp}, \&
  {Launhardt}}]{Mez99}
{Mezger}, P.~G., {Zylka}, R., {Philipp}, S., \& {Launhardt}, R. 1999, \aap,
  348, 457

\bibitem[{{Miralda-Escud{\' e}} \& {Gould}(2000)}]{Mir00}
{Miralda-Escud{\' e}}, J. \& {Gould}, A. 2000, \apj, 545, 847

\bibitem[{{Morris}(1993)}]{Mor93}
{Morris}, M. 1993, \apj, 408, 496

\bibitem[{{Mouawad} {et~al.}(2004){Mouawad}, {Eckart}, {Pfalzner}, {Sch\"odel},
  {Moultaka}, \& {Spurzem}}]{Mou04}
{Mouawad}, N., {Eckart}, A., {Pfalzner}, S., {Sch\"odel}, R., {Moultaka}, J.,
  \& {Spurzem}, R. 2004, {{\aap} submitted; available at
  http://xxx.lanl.gov/abs/astro-ph/0402338}

\bibitem[{{Portegies Zwart} {et~al.}(2002){Portegies Zwart}, {Makino},
  {McMillan}, \& {Hut}}]{Por02}
{Portegies Zwart}, S.~F., {Makino}, J., {McMillan}, S.~L.~W., \& {Hut}, P.
  2002, \apj, 565, 265

\bibitem[{{Portegies Zwart} {et~al.}(2003){Portegies Zwart}, {McMillan}, \&
  {Gerhard}}]{Por03}
{Portegies Zwart}, S.~F., {McMillan}, S.~L.~W., \& {Gerhard}, O. 2003, \apj,
  593, 352

\bibitem[{{Sch{\" o}del} {et~al.}(2003){Sch{\" o}del}, {Ott}, {Genzel},
  {Eckart}, {Mouawad}, \& {Alexander}}]{Sch03}
{Sch{\" o}del}, R., {Ott}, T., {Genzel}, R., {Eckart}, A., {Mouawad}, N., \&
  {Alexander}, T. 2003, \apj, 596, 1015

\bibitem[{{Schaller} {et~al.}(1992){Schaller}, {Schaerer}, {Meynet}, \&
  {Maeder}}]{Sch92}
{Schaller}, G., {Schaerer}, D., {Meynet}, G., \& {Maeder}, A. 1992, \aaps, 96,
  269

\bibitem[{{Syer} \& {Ulmer}(1999)}]{Sye99}
{Syer}, D. \& {Ulmer}, A. 1999, \mnras, 306, 35

\bibitem[{{Vollmer} \& {Duschl}(2001)}]{Vol01}
{Vollmer}, B. \& {Duschl}, W.~J. 2001, \aap, 377, 1016

\bibitem[{{Zhao} {et~al.}(2002){Zhao}, {Haehnelt}, \& {Rees}}]{Zha02}
{Zhao}, H., {Haehnelt}, M.~G., \& {Rees}, M.~J. 2002, New Astronomy, 7, 385

\end{thebibliography}

\end{document}